\documentclass[12pt]{article}
\usepackage{pslatex}
\usepackage{txfonts}
\usepackage{latexsym}
\usepackage{graphicx}
\usepackage{epsfig}

\begin{document}

\author{V. H. Hamity\footnote{\texttt{hamity@famaf.unc.edu.ar}}, M. A. C\'ecere\footnote{\texttt{cm4@famaf.unc.edu.ar}}
 and D.E. Barraco\footnote{\texttt{barraco@famaf.unc.edu.ar}}\\ Fa.M.A.F., Universidad Nacional de C\'ordoba\\
  Ciudad Universitaria, C\'ordoba 5000, Argentina}

\title{Relativistic dynamics of cylindrical shells of counter-rotating particles}
\maketitle

\begin{abstract}
Although infinite cylinders are not astrophysical entities, it is
possible to learn a great deal about the basic qualitative features
of generation of gravitational waves and the behavior of the matter
conforming such shells in the limits of very small radius. We
describe the analytical model using kinetic theory for the matter
and the junction conditions through the shell to obtain its equation
of motion. The nature of the static solutions are analyzed, both for
a single shell as well as for two concentric shells. In this second
case, for a time dependent external shell, we integrate numerically
the equation of motion for several values of the constants of the
system. Also, a brief des\-crip\-tion in terms of the Komar mass is
given to account for the gravitational wave energy emitted by the
system.

\end{abstract}

\section{Introduction}
Recently there has been a growing interest in the literature to
study the gravitational dynamics of infinite cylindrical shells of
counter-rotating particles in different geometrical
settings\cite{trabajosprevios}. One of the best studied case is that
considered by Apostolatos and Thorne\cite{AT}. They made a detailed
analysis of the system using C-energy balance arguments\cite{thorne}
and a sequence of momentarily static and radiation-free shells to
avoid the complications causes by gravitational radiation for a
fully relativistic cylindrical shell. However, for a shell outside
equilibrium the radiation-free requirement is incompatible with the
shell equation of motion, obtained from the matching conditions  of
the interior and exterior vacuum solutions. In this paper we show
that there are two equilibrium solutions, corresponding to constants
values of shell radius, the rest mass per unit of proper length
($\Lambda$) and the angular momentum per unit of rest mass ($J$) of
the particles composing the shell. One of the solution is a stable
equilibrium while the other is unstable. Next, we study a system of
two shells, first in a static configuration and show that the
equilibrium configuration of the external shell may also be stable
or unstable in a similar way as for only one shell; then, we
integrate (numerically) the equation of motion of the external shell
for different values of the constants of motion and initial values
of the coordinate radius starting at rest. This integration is
possible, without having a complete knowledge of the wave solution
outside the external shell, because the matching conditions and the
fact that the internal solutions sa\-tis\-fies the time independent
field equations. The innermost shell is chosen in a static stable
configuration, which allows us to have a smooth spacetime geometry
on the cylinder's symmetry axis. We obtain shells that perform
damped oscillations, collapse or are  locally expanding, depending
on the values of the conserved quantities and integration constants.
We give a brief description of the energy emitted by the system in
the case of damped oscillations, in terms of a generalized Komar
mass.

The paper is organized as follows. In the next section we describe
the geo\-metry and  equation of motion of a cylindrical shell of
counter-rotating particles. This includes the metric and junction
conditions and the description of the surface stress-energy tensor
using the kinetic theory in general Relativity adapted to the shell.
In section 3 we discuss the stable and unstable  equilibrium
solutions of a single shell. In section 4 the dynamics of two
concentric cylindrical shells of counter-rotating particles is
presented and the matter equation of motion is numerically
integrated to show the different types of solutions, depending on
the values of the conserved quantities and integration constants.
Finally, a brief summary of the main results is given in section 5.

\section{Geometry and equation of motion of a cylindrical shell of counter-rotating particles}
\subsection{Metric and junction conditions}

We consider a spacetime $M=M^-\cup \Sigma \cup M^+$ with cylindrical
symmetry where $\Sigma$ is the history of a hollow cylinder composed
of counter-rotating particles of rest mass equal to unity; $M^-$
($M^+$) is the vacuum interior (exterior) region of the cylinder. In
the vacuum interior ($M^-$) and exterior ($M^+$) of the shell, we
introduce canonical cylindrical coordinates $(t,r,z,\phi)$. The
metric takes the form\cite{thorne}.
\begin{equation}\label{1}
    ds_{\pm}^{2}=e^{2\gamma_{\pm}-2\psi_{\pm}}(dr^{2}-dt_{\pm}^{2})
                 +e^{2\psi_{\pm}}dz^{2}+e^{-2\psi_{\pm}}r^{2}d\phi^{2}.
\end{equation}

These coordinates are uniquely determined up to the (non-trivial)
change of scale
\begin{equation}\label{2}
z \longrightarrow e^\mu z, \ \ \ \ \psi \longrightarrow \psi - \mu,
\ \ \ \ r \longrightarrow e^{-\mu} r, \ \ \ \ t \longrightarrow
e^{-\mu} t.
\end{equation}

The Einstein field equations in the empty space inside and outside
the shell are

\begin{equation}\label{3}
\psi_{,rr}+\frac{1}{r}\psi_{,r}-\psi_{,tt}=0 \\
\end{equation}
\begin{equation}\label{4}
\gamma_{,t}=2r\psi_{,r}\psi_{,t}, \ \ \ \
\gamma_{,r}=r(\psi_{,r}^{2}+\psi_{,t}^{2}).
\end{equation}

Thus, $\psi(r,t)$ plays the role of a gravitational field whose
static part is analogue of the Newtonian potential. The time
depended solutions of (\ref{3}) represent gravitational
waves\cite{Einstein-Rosen}. Equation (\ref{3}) is the integrability
condition of equations (\ref{4}). The coordinates  $(z,\phi,r)$ and
the metric function $\psi$ are continuous across the shell $\Sigma$,
while $t$ and the metric function $\gamma$ are discontinuous.
Smoothness of the spacetime geometry on the axis $r=0$ requires that
\begin{equation}\label{5}
\gamma=0 \;\;\; \mbox{and}\;\;\;  \psi \;\;\; \mbox{finite at}
\;\;\; r=0.
\end{equation}

The junction conditions of $M^-$ and $M^+$ through $\Sigma$ require
the continuity of the metric and to specify the jump of the
extrinsic curvature $K^{\pm}$ compatible with the stress-energy
tensor on the shell. The induced metric on $\Sigma$ is given by
\begin{equation}\label{6}
ds^{2}_{\Sigma}=-d\tau^{2}+e^{2\psi_{\Sigma}}dz^{2}+e^{-2\psi_{\Sigma}}R^{2}d\phi^{2}.
\end{equation}

Here
$\psi_{\Sigma}(\tau)=\psi_{+}(R(\tau),t_{+}(\tau))=\psi_{-}(R(\tau),t_{-}(\tau))$;
the evolution of the shell is characterized by $R(\tau)$, which is
the radial coordinate $r$ at the shell's location and $\tau$ the
proper time of an observer at rest on $\Sigma$. The Einstein field
equations on the shell reduce to\cite{Israel}
\begin{equation}\label{7}
K_{ij}^{+}-K_{ij}^{-}=8\pi(S_{ij}-\frac{1}{2}S g_{ij});
\end{equation}
here $S_{ij}$ es the surface stress-energy tensor, $g_{ij}$ is given
by (\ref{6}) $[i=(\tau,z,\phi)]$ and

\begin{equation}\label{8}
K_{ij}^{\pm}=-(h^{c}_{d}h^{a}_{b}N^{\pm}_{c;a})e^{d}_{i}e^{b}_{j}\;;
\ \ \ \
e_{i}=(\frac{\partial}{\partial\tau},\frac{\partial}{\partial
z},\frac{\partial}{\partial\phi})\;; \ \ \ \
h^{c}_{d}=\delta^{c}_{d}-N_{d}N^{c},
\end{equation}
$N$ is the outward unit vector normal to the shell:
\begin{equation}\label{8'}
N=N^r \frac{\partial}{\partial r}+N^t \frac{\partial}{\partial t};
\;\;\; N^r>0\;.
\end{equation}
The tangent vector is
\begin{equation}\label{9}
\frac{\partial}{\partial\tau}=\dot{R}\frac{\partial}{\partial
r}+X^{\pm}\frac{\partial}{\partial t^{\pm}}\;\;.
\end{equation}
From the normalization conditions $(u \cdot u=-1 \,,\, N \cdot N =
1)$ we get
\begin{equation}\label{10}
\frac{\partial t_{\pm}}{\partial\tau}\equiv X^{\pm}= +
\sqrt{e^{-2(\gamma_{\pm}-\psi_{\Sigma})}+\dot{R}^{2}},
\end{equation}
Similarly, from $u \cdot N=0$, we get
\begin{equation}\label{11}
N^r=X^{\pm}, \ \ \ \ \ \ N^t=\dot{R}.
\end{equation}
The calculation of the tensors $K_{ij}^{\pm}$ gives the result (we
omit the $(\pm)$ for simplicity)
\begin{eqnarray}
K_{\tau\tau}&=&(\gamma_{,t}-\psi_{,t})\dot{R}+X(\gamma_{,r}-\psi_{,r})-(\dot{R}\dot{X}-\ddot{R}X)e^{2\gamma-2\psi_{\Sigma}}, \label{12}\\
K_{\phi\phi}&=&-R^{2}e^{-2\psi_{\Sigma}}\biggl[\left(\frac{1}{R}-\psi_{,r}\right)X-\psi_{,t}\dot{R}\biggr], \label{13}\\
K_{zz}&=&-e^{2\psi_{\Sigma}}[\psi_{,r}X+\psi_{,t}\dot{R}].\label{14}
\end{eqnarray}
All the quantities are computed at the location of the shell
$\Sigma$.

\subsection{The surface  stress-energy tensor}

To compute the surface stress-energy tensor we shall use the
relativistic kinetic theory\cite{Ehlers} adapted to the matter
within the shell
$\Sigma$. \\
The stress-energy tensor at the event $x \in \Sigma$ is defined in
the usual way by
\begin{eqnarray}\label{15}
T^{ab}(x)&=&\int_{P(x)}p^{a}p^{b}f(x,p)\ \pi \nonumber \\
&=&S^{ab}\delta(n),
\end{eqnarray}
here $f(x,p)$ is the distribution function; $P(x)$ is the space of
momentum $p = ( p^a \partial / \partial x^a)$ in $x$, corresponding
to particles with proper mass $ m = 1$, $p \cdot p = - 1$, and $\pi$
is the intrinsic element of volume in this space. Introducing  a
coordinate $n$ normal to $\Sigma$, such that $n = 0$ defines
$\Sigma$,  we have $x^a = (n, x^i)$ and
$$S^{an}=S^{na}=0,$$
\begin{equation}\label{16}
S^{ij}=\int_{\tilde{P}(x)} p^{i}p^{j} \tilde{f} \ \tilde\pi \,,
\end{equation}
$$\tilde\pi =\frac{\sqrt{-g}}{|p_\tau|}dp^z dp^\phi, \ \ \ \ \ \ g=det(g_{ij})\;\;. $$
Here $\tilde P(x)=\{p:p \in P(x),p \cdot n=0\}$; $\tilde f$ is the
restriction to $\Sigma$ of the distribution function $f$. In absence
of collisions, $\tilde f$ satisfies the Liouville equation in the
one-particle phase space $\tilde{M}=\{(x,p)$ / $x \in \Sigma$; $p
\in \tilde{P}(x)$\}.
\begin{equation}\label{17}
    \tilde{L}(\tilde f)\equiv \frac{d\tilde{f}}{d\tau} = 0\;\;,
\end{equation}
where $(d\tilde{f}/d\tau)$ means the derivative of $\tilde{f} (x^i,
p^j)$ along a phase orbit. Since\cite{Ehlers}
\begin{equation}\label{18}
    S^{ij}_{;j}=\int_{\tilde P}{p^i \tilde{L}(\tilde f)}
    \tilde{\pi}\;\;,
\end{equation}
we get from (\ref{17}) and (\ref{18}) the conservation law
$S^{ij}_{;j}=0$. Similarly the particle four-current density,
defined by
\begin{equation}\label{19}
    {\cal N}^{i}=\int_{\tilde{P}(x)}\tilde{f} p^{i} \tilde{\pi}\;\;,
\end{equation}
verifies
\begin{equation}\label{20}
{\cal N}^i_{;i}=\int_{\tilde{P}(x)}\tilde{L}(\tilde
f)\tilde{\pi}\;\;.
\end{equation}
A simple way of constructing a solution of the Liouville equation is
to consider $\tilde f$ as a function of constants of the motion. In
our case, due to the cylindrical symmetry of the source, the
components $p_z$ and $p_{\phi}$ of the canonical momentum are
conserved. The component $p_{\phi}$ is the particle angular
momentum. Thus, a solution, $\tilde f (p_z,p_{\phi})$, of Liouville
equation, for particles with $p_z=0$, and counter-rotating is given
by
\begin{equation}\label{21}
    \tilde f (p_z,p_{\phi})=k\delta(p_z)\delta(p^2_{\phi}-J^2)\;\;.
\end{equation}
Here $J$ is the modulus of angular momentum per unit mass. We have
the same number of particles with $p_{\phi}=J$ and $p_{\phi}=-J$;
$k$ is a constant related to the density of particles at a given
point on $\Sigma$.

The expression (\ref{16}) can be written in the form:
\begin{equation}\label{22}
    S^{ij}=\int_{\tilde P(x)}{p^i p^j e^{-2 \psi_{\Sigma}}\delta(p^z)\frac{k}{2J}[\delta(p_{\phi}-J)+
    \delta(p_{\phi}+J)]\frac{\sqrt{-g}}{|p_{\tau}|}dp^{z}dp^{\phi}}\;\;.
\end{equation}
To compute $S^{ij}$ we have to take into account that
$p^\phi=g^{\phi\phi}p_{\phi}$. The non-zero components of $S^{ij}$
are:
\begin{eqnarray}
    S^{\phi\phi}&=&\frac{kJe^{4\psi_{\Sigma}}}{ER^5}\equiv pg^{\phi\phi}\;\;, \label{23}\\
    S^{\tau\tau}&=&\frac{kE}{JR}\equiv \eta \;\;.\label{24}
\end{eqnarray}
The quantities $\eta$ and $p$ are the surface energy density and
pressure. The shell's full stress-tensor is
\begin{equation}\label{25}
S=\eta \ u \otimes u + p (g + u \otimes u - \upsilon \otimes
\upsilon)\;\;.
\end{equation}
Here $\upsilon=e^{-\psi_{\Sigma}}e_z$ is the unit vector in the $z$
direction; $E=-p \cdot u$. \\
From $p \cdot p=- 1$ we have:
\begin{equation}\label{26}
\frac {e^{2\psi_{\Sigma}}J^{2}}{R^{2}}=E^{2} - 1 \equiv w^2,
\end{equation}
where $w$ is the modulus of the particle linear momentum per unit
mass. Introducing the circumference of the shell $2\pi {\cal R} =2
\pi R e^{-\psi_{\Sigma}}$, we have
\begin{equation}\label{27}
    w=\frac{e^{\psi_{\Sigma}}J}{R}=\frac{J}{{\cal R}}\;\;.
\end{equation}
Thus
\begin{eqnarray}
\eta &=&\frac{kE}{JR}\ \ \ , \label{28}\\
 p&=&\eta \left( \frac{E^2 -
1}{E^2} \right)\;\;.  \label{28'}
\end{eqnarray}
Using (\ref{19}) and (\ref{21}) we have
\begin{equation}\label{29}
    {\cal N}=\nu u\;\;,
\end{equation}
where $\nu$ is the (surface) number of particles density.
Integration of (\ref{19}) gives
\begin{equation}\label{30}
    \nu = \frac{k}{JR} \ \ \ \ \Rightarrow \ \ \ \ 2\pi\nu R=\frac{2\pi
    k}{J}\equiv \lambda \;\;,
\end{equation}
where $\lambda \equiv (dm /dz)$ is a conserved quantity: the shell's
total rest mass per unit Killing length $z$\cite{AT}. The shell's
total rest mass per unit of proper length ($dz e^{\psi}=1$) is :
\begin{equation}\label{31}
    \Lambda \equiv -\int{{\cal N}^i u_i R e^{-\psi}d\phi}=\lambda
    e^{-\psi}\;\;.
\end{equation}
From (\ref{28}) and (\ref{30}) we have $\eta=\nu E$. Summing up, the
parameters ${\cal R}, \Lambda, w, E, p, J$ are invariant by the
rescaling (\ref{2}). On the other hand, $R$ y $\lambda$ are scale
dependent. The quantity $\lambda$ and $J$ are conserved quantities
during a time evolution of the shell.

\subsection{ The shell's equation of motion}

From equations (\ref{12},\ref{13},\ref{14}) and (\ref{25}) the
junction
conditions (\ref{7}) become: \\
The $zz$ component:
\begin{equation}\label{32}
    \psi^{+}_{,n}-\psi^{-}_{,n}=4 \pi (p-\eta)\;\;.
\end{equation}
The $\phi\phi$ component:
\begin{equation}\label{33}
    -\frac{X^+ - X^-}{R}+\psi^{+}_{,n}-\psi^{-}_{,n}=4 \pi
    (p+\eta)\;\;.
\end{equation}
The $\tau\tau$ component:
\begin{equation}\label{34}
    (\gamma -  \psi)^{+}_{,n}-(\gamma -  \psi)^{-}_{,n}-(A^+ -
    A^-)=4\pi (\eta +p)\;\;.
\end{equation}
Here $A\equiv e^{2(\gamma-\psi)}(\dot{R}\dot{X}-\ddot{R}X)$;
$\psi_{,n}\equiv N(\psi)$ is the normal derivative of the
gravitational wave field. Equations (\ref{32}) and (\ref{34}) are
equivalent to:
\begin{eqnarray}
\psi^{+}_{,n}-\psi^{-}_{,n}=-\frac{2\lambda}{\sqrt{1+w^2}R}\;\;, \label{35}\\
 X^+ - X^-=-4\lambda \sqrt{1+w^2}\;\;. \label{36}
\end{eqnarray}
Using the vacuum field equations (\ref{3}) and (\ref{4}), and the
junction conditions (\ref{35}) and (\ref{36}), equation (\ref{34})
becomes:
\begin{equation}\label{37}
\frac{d^{2}R}{d\tau^2}=\dot{R}\dot{\psi_{\Sigma}}-R[(\dot{\psi_{\Sigma}})^2+(\psi^{-}_{,n})^2]
+\frac{\psi^{-}_{,n}X^{-}}{1+w^2}-\frac{\lambda
X^{-}}{R(1+w^2)^\frac{3}{2}}+\frac{w^2X^{-}X^{+}}{R(1+w^2)}\;.
\end{equation}
Equations (\ref{35}-\ref{37}) are the same as (24.a-b-c) of
Apostolato and Thorne\cite{AT}.
\section{Static solutions}
\subsection{The metric}

For a static configuration we have $\dot{R}=0$; $\ddot{R}=0$,
$\psi_{,t}=\psi_{,tt}=0$. Then, in the vacuum outside the shell the
field equations (\ref{3}) and (\ref{4}) imply
\begin{eqnarray}
    \psi^{+}(r)&=&\psi_{\Sigma}-\kappa \,ln(r/R) \ \ \ \ \ \ \
    r > R\;\;, \label{38}\\
    \gamma^{+}(r)&=&\gamma+\kappa^{2} \,ln(r/R) \ \ \ \ \ \ \
    r > R \;\;,\label{39}
\end{eqnarray}
where $\kappa,\, \psi_{\Sigma}$ and $\gamma$ are constants.
\begin{figure}
\centering \epsfxsize=8.5cm \epsfbox{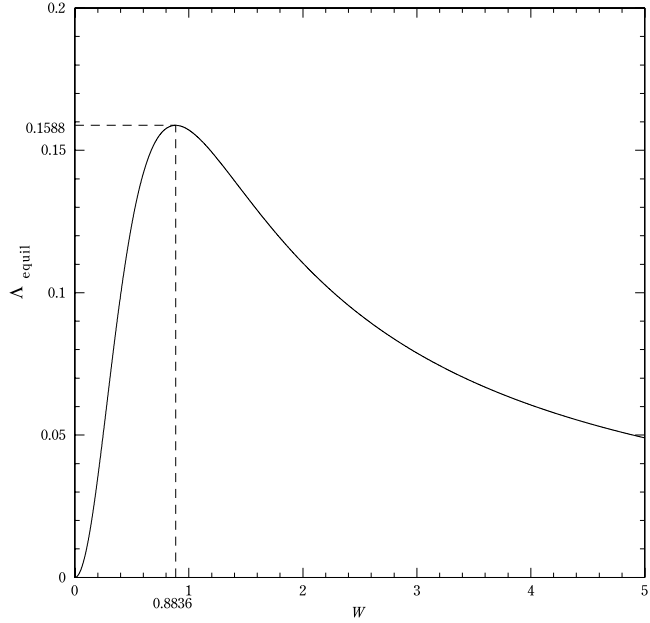} \vspace{-0.3cm}
\caption{\small{Rest mass per unit proper length for the equilibrium
configurations versus the linear momentum per unit rest mass of the
particles [equation (\ref{44})]. For $w < 0.8836$ the equilibrium is
stable while for $w
> 0.8836$ unstable .}} \label{A}
\end{figure}
Similarly, in the vacuum inside the shell the field equations
(\ref{3}) and (\ref{4}) plus the boundary conditions (\ref{5}) give
\begin{equation}\label{40}
    \psi^- =\psi_{\Sigma}; \ \ \ \ \ \gamma^- = 0 \ \ \ \ \ \ \
    r<R\;\;.
\end{equation}
From (\ref{10}) we have
\begin{equation}\label{41}
    X^+=e^{\psi_{\Sigma}-\gamma}; \ \ \ \ \ \
    X^-=e^{\psi_{\Sigma}}\;\;.
\end{equation}
Then, from (\ref{36}) and (\ref{41}) we obtain
\begin{equation}\label{42}
    e^{-\gamma}=1-4\Lambda\sqrt{1+w^2}\;\;.
\end{equation}
Using that $\dot{R}=0, \,\dot{\psi}_{\Sigma}=0, \,\psi^{-}_{,n} = 0$
and (\ref{42}), the equation of motion (\ref{37}) takes the form:
\begin{equation}\label{43}
\ddot{\cal{R}}\equiv e^{-\psi}\ddot{R}= \frac{(2w^2+1)^2}{{\cal
R}(1+w^2)^{3/2}}[\Lambda_{eq}-\Lambda]\;\;,
\end{equation}
where
\begin{equation}\label{44}
    \Lambda_{equil}(w) = \frac{w^2\sqrt{1+w^2}}{(2w^2+1)^2}\;\;.
\end{equation}
In the case of an equilibrium configuration of the shell, we have:
\begin{equation}\label{45}
    \ddot{R}=0 \ \ \ \ \ \ \ \Leftrightarrow \ \ \ \ \ \ \
    \Lambda = \Lambda_{equil}(w)\;\;.
\end{equation}
In figure (\ref{A}) we plot $\Lambda_{equil} (w)$; the maximum value
$\Lambda_{equil,max}=0.1588$ is reached for $w=0.8836$. Given a
constant $\Lambda < 0.1588$ we find from (\ref{44}) two values of
$w$ that correspond to equilibrium configurations; or, we may use
this equation to obtain a value of $\Lambda$ from $w$.  Recall that
$w=J/{\cal R}$; i.e., for fix $J,\, {\cal R}$ decreases while $w$
increases. For a fix value of $w$ (or $\Lambda$), $\cal R$ is
proportional to $J$. Thus, we may have $\cal R$ approaching zero by
just taking $J$ very small.

From the junction condition (\ref{35}) and equations
(\ref{38}-\ref{42}), (\ref{44}) and (\ref{45}), we obtain
\begin{eqnarray}
    \kappa&=&2w^2\;\;;     \label{46} \\
    \gamma&=&ln(1+2w^2)^2  \;\;.  \label{47}
\end{eqnarray}
Thus, all the scale invariant quantities that define the static
configuration are determined; the scale dependent constant
$\psi_{\Sigma}$ may be computed from the relations
$e^{-\psi_{\Sigma}}=\Lambda / \lambda={\cal R}/R$, if a value of
$\lambda$ or $R$ is given (fixing the coordinate system). In the
next subsection we analyze the stability of the equilibrium
configurations.
\begin{figure}
\centering \epsfxsize=8.5cm \epsfbox{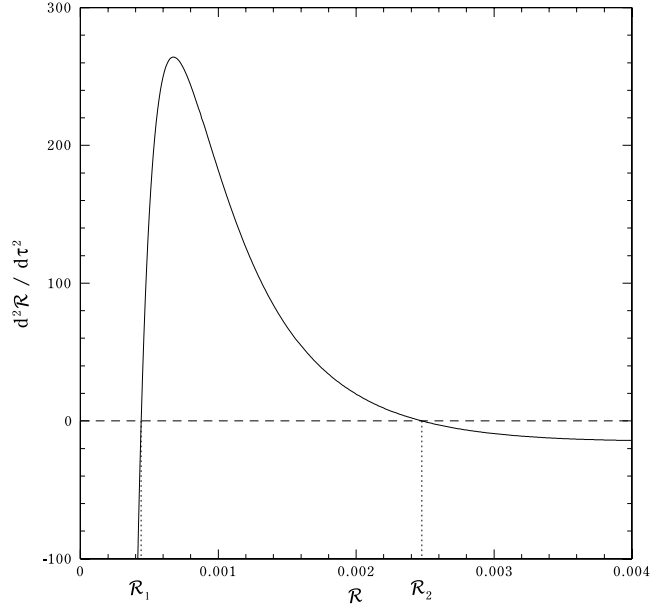} \vspace{-0.3cm}
\caption{\small{Acceleration vs proper radius of the shell for a
momentarily static and radiation free shell\cite{AT}  given by
(\ref{43}). The shell with radius ${\cal R}_1 = 0.00044$ is unstable
while ${\cal R}_2 = 0.0024$ corresponds to a static stable shell. }}
\label{B}
\end{figure}

\subsection{Stability analysis}

To determine the nature of the static configurations described in
the previous subsection we perform a virtual displacement $\delta
{\cal R}$ from its equilibrium value. Then,  we compute from
(\ref{43}) the virtual change in the acceleration $\ddot{\cal R}$,
around $\ddot{\cal R}=0$ in the form
\begin{equation}\label{48}
    \delta \ddot{{\cal R}}=\frac{\partial \ddot{{\cal R}}}{\partial {\cal
    R}}  {\bigg\vert_{{\cal R}_{eq}}}      \delta {\cal R}\;\;.
\end{equation}
The shell is in a static stable configuration if $(\partial
\ddot{{\cal R}}/ \partial {\cal R}) | _{{\cal R}_{eq}}<0$; similarly
the equilibrium configuration is unstable if $(\partial \ddot{{\cal
R}}/
\partial {\cal R}) | _{{\cal R}_{eq}}>0$. From (\ref{43}), we
have
\begin{equation}\label{49}
    \frac{\partial \ddot{{\cal R}}}{\partial {\cal R}}{\bigg\vert_{{\cal
    R}_{eq}}}=-{positive\choose quantity}\times \frac{d \Lambda_{equil}}{d
    w}\;\;.
\end{equation}
From figure (\ref{A}) we immediately conclude that the stable static
configurations co\-rres\-pond to  $w<0.8836$; i.e., for values of
${\cal R}>J/0.8836$ (this includes the Newtonian limit); while for
$w>0.8836$, ${\cal R}<J/0.8836$, the equilibrium configurations are
unstable. In figure (\ref{B}) we plot $\ddot{{\cal R}}$ as a
function of ${\cal R}$,  for $J=0.001$; $\Lambda=0.1$. Notice that
at ${\cal R}_1=0.00044$, $(\partial \ddot{{\cal R}}/
\partial {\cal R}) | _{{\cal R}_{1}}>0$ it is sensible larger than
$\mid\partial \ddot{{\cal R}}/\partial {\cal R}\mid_{{\cal R}_{2}}$
for ${\cal R}_2=0.0024$;  ${\cal R}_1$ and ${\cal R}_2$ are the
unstable and stable radius, respectively, of the shell.

\section{Dynamics of two cylindrical shells of counter-rotating particles}

\subsection{The model}

We consider two concentric hollow cylinders composed of
counter-rotating particles which define singular hypersurfaces
$\Sigma$ and $\Sigma'$ of  coordinate radius $R$ and $R'$,
respectively ($R'>R$), such that the spacetime $M=M^{0}\cup \Sigma
\cup M^{-} \cup \Sigma' \cup M^{+}$. In the vacuum interior ($M^0$),
intermediate region ($M^-$), and exterior ($M^+$), we use canonical
cylindrical coordinates $(t,r,z,\phi)$ as before (see (\ref{1})).
The hypersurface $\Sigma$ is considered in a static, stable
equilibrium ($R$=const). Therefore the region $M^0$ and $M^-$ are
related through $\Sigma$ in the same way as in section 3. In
general, the radius of the singular hypersurface $\Sigma'$ will be
time dependent. Thus, the metric coefficients in the different
regions are given by:
\begin{eqnarray}
&a)& 0< r <R : \nonumber\\
& & \psi_0=\psi_{\Sigma} = \mbox{constant}\;\;;\;\; \gamma_0=0\;.
 \label{50}\\
& b)& R<r<R'(\tau): \nonumber \\
& &\psi^{-}(r) =-\kappa \ ln(r/R) + \psi_{\Sigma}\;\; ; \;\;
 \label{51} \\
& &\gamma^{-}(r)=\kappa ^2 \ ln(r/R) + \gamma  \label{52} \\
& &\kappa=2 \, w^2\;; \;\; \gamma=ln(1+2w^2)^2\;. \nonumber \\
&c)& R'(\tau)<r : \nonumber \\
& & \psi^+ (r,t^+)\;\;;\;\;\gamma^+(r,t^+) \;\;\;(\mbox{wave
solutions of the field equations}) \nonumber
\end{eqnarray}
The matching conditions at $r_{\Sigma '}=R'(\tau)$  require
\begin{eqnarray}
& &\psi^{+}(R'(\tau),t^{+}(\tau))=\psi^{-}(R'(\tau))\equiv
\psi_{\Sigma'}(\tau)\;\;,
\nonumber \\
& &\psi_{\Sigma'}(\tau)=-\kappa \ ln(R'(\tau)/R)+\psi_{\Sigma}\;\;.
\label{53}
\end{eqnarray}
The shells' surface stress-energy tensors are (see subsection 2.2)
\begin{eqnarray}
S&=&\eta \ u \otimes u + p (g + u \otimes u - \upsilon \otimes \upsilon)\;\;;\label{54} \\
S'&=&\eta' \ u' \otimes u' + p' (g' + u' \otimes u' - \upsilon'
\otimes \upsilon')\;\;, \label{55}
\end{eqnarray}
with obvious meanings for the symbols.
\subsection{Static solutions}
\begin{figure}
\centering \epsfxsize=7.5cm \epsfbox{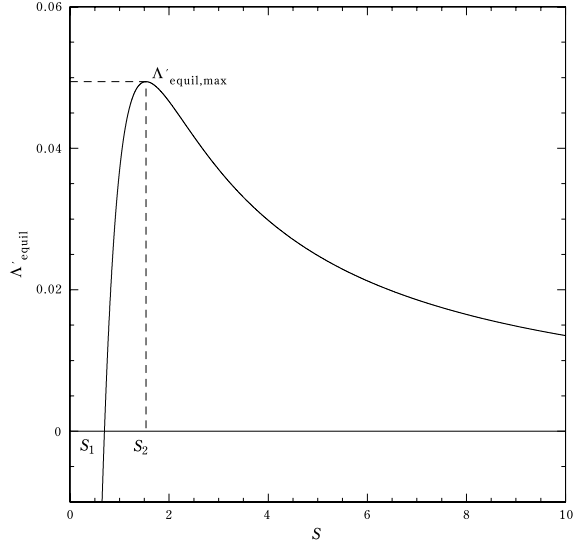} \vspace{-0.3cm}
\caption{\small{Rest mass per unit proper length of equilibrium
configurations of the external shell  versus linear momentum per
unit rest mass of the particles [equation(\ref{59})]. The stable
configurations correspond to $s_1 < s < s_2$. For $s > s_2$ the
equilibrium is unstable; $s_1 = 0.6957\,,  s_2 = 1.5346$.}}
\label{C}
\end{figure}
A static solution corresponds to $R'=const$. The metric coefficients
in $M^+$ are
\begin{eqnarray}
    \psi^{+}(r)&=&-\kappa' \ ln(r/R')+\psi_{\Sigma'} \;\;, \label{56}\\
    \gamma^{+}(r)&=&-\kappa'^2 \ ln(r/R')+\gamma'' \;\;, \label{57}
\end{eqnarray}
\begin{figure}
\centering \epsfxsize=7.5cm \epsfbox{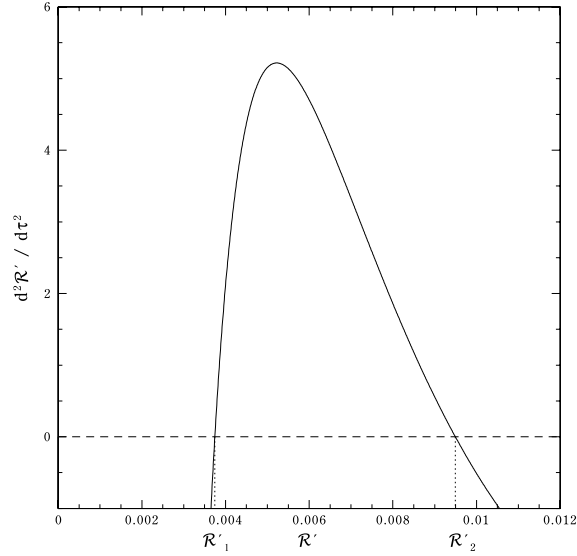} \vspace{-0.3cm}
\caption{\small{Acceleration vs proper radius of a momentarily
static   and radiation free external shell\cite{AT}. The plot
correspond to equation (\ref{61}). The shell with radius ${\cal
R'}_1 = 0.003744$ is unstable while ${\cal R'}_2 = 0.009493$
corresponds to a static stable shell. }} \label{D}
\end{figure}
where $\kappa',\,\psi_{\Sigma'}\,,\gamma''$are constants to be
determined. Equations (\ref{36}) and (\ref{37})
($\dot{R'}=0=\ddot{R'}$) become
\begin{equation}\label{58}
(e^{-\gamma^+}-e^{-\gamma^-})|_{R'}=-4\Lambda'E'\, ,
\end{equation}
\begin{equation}\label{59}
\Lambda'=\frac{\sqrt{1+s^2}[s^2-2w^2(1+s^2)]}{(1+2s^2)^2(1+2w^2)}\left(\frac{\cal
R}{\cal R'}\right)^{\frac{4w^4}{1 + 2 w^2}}\equiv
\Lambda'_{equil}(s,w, R, J')\;\;.
\end{equation}
Where
\begin{equation}\label{60}
    E'=\sqrt{1+s^2}, \ \  s=\frac{J'}{{\cal R'}}, \ \ \gamma^+ (R')
    = \gamma''
    \;, \ \ \gamma^- (R')
    = \gamma + ln (R'/R)^{\kappa^2}\equiv \gamma'\;\;,
\end{equation}
$J'$ is the angular momentum per unit mass of the particles
constituting  $\Sigma'$, ${\cal R'}=R'e^{-\psi_{\Sigma'}}$. Notice
that (\ref{59}) reduces to (\ref{44}) for $w = 0$. From (\ref{53})
we have
\begin{equation}\label{62}
    \frac{{\cal R'}}{{\cal
    R}}=\left(\frac{R'}{R}\right)^{\kappa+1}\;\;.
\end{equation}
In figure (\ref{C}) we show $\Lambda'_{equil}(s,w, R, J')$ as a
function of $s$ for $w = 0.4038$,  $R = 1$ (the values corresponding
to the stable static shell of section 3) and $J' = 0.01 $.
Considering $\dot{R}=0, \ \dot{\psi}_{\Sigma'}=0, \
\psi^{-}_{,n}|_{R'}= -(2w^2/{\cal R'})e^{-\gamma'}$, a similar
calculation as the one done for the interior shell ($\Sigma$), shows
that (\ref{37}) becomes
\begin{equation}\label{61}
    \ddot{{\cal R'}}= \left(\frac{\cal
R}{\cal R'}\right)^{\frac{4w^4}{1 + 2 w^2}} \frac{(1+2s^2)^2}{{\cal
R'}(1 + 2 w^2)^2 (1+s^2)^{3/2}}[\Lambda'_{equil}-\Lambda']\;\;.
\end{equation}
Thus, the static stable configuration of the exterior shell
($\Sigma'$) correspond to those values of $s$, such that $(\partial
\Lambda'_{equil}/\partial s)>0$. For $w = 0.4038$$,  {\cal R} =
0.002476$, $J'= 0.01$ we have a stable configuration for $ 0.6957 <
s < 1.5346$ [see  figures (\ref{C} and \ref{D})\,;\,$s = 1.05335$].

To complete the static solution we need the expressions for
$\kappa'$ and $\gamma''$. A lengthy but straight forward calculation
gives:
\begin{eqnarray}
\kappa' &=& \frac{2 s^2 + 2 w^2 (1 + s^2) (4 s^2 - 2 w^2)}{1 + 8 w^2
(1 + s^2) + 8 w^2 (1 + s^2)^2} \;\;, \label{61'}\\
e^{-\gamma''} &=& \left(\frac{\cal R}{\cal R'}\right)^{\frac{4w^4}{1
+ 2 w^2}}\frac{1 + 8 w^2 (1 + s^2) [ s^2 (1 + 2 w^2) + 2 w^2]}{(1 +
2 w^2)^2 (1 + 2 s^2)^2} \;\;. \label{62"}
\end{eqnarray}

\subsection{Damped oscillating solutions}
\begin{figure}
\centering \epsfxsize=8.5cm \epsfbox{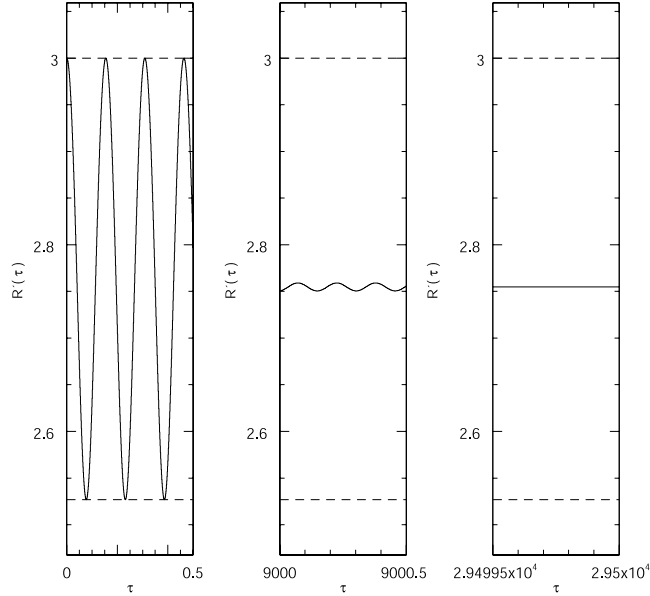} \vspace{-0.3cm}
\caption{\small{Sequence $R'(\tau)$ of a damped oscillating solution
for $\lambda'= 11.6070, \ \ J'=0.01, \ \ R'_i = 3$.}} \label{E}
\end{figure}
Let us now consider a simple time dependent model by assuming that
the external shell is time dependent: $r_{\Sigma'}=R'(\tau)$. We can
integrate the equation of motion (\ref{37}) for ${\cal R'}(\tau)$
without knowing explicitly the metric for $r>R'(\tau)$. Outside the
external shell we have  a wave solution of the field equations
(\ref{3}, \ref{4}) that satisfies the boundary conditions
\begin{eqnarray}
    \psi^{+}(R'(\tau),t^+(\tau))=\psi_{\Sigma'}(\tau)\;, \label{63} \\
    \gamma^{+}(R'(\tau),t^+(\tau))=\gamma''(\tau)\;, \label{64}
\end{eqnarray}
where $\psi_{\Sigma'}(\tau)$ is given by (\ref{53});
$\gamma''(\tau)$ may be obtained from the matching condition
(\ref{36})  at the hypersurface $\Sigma'$. We specify a particular
system choosing an internal static stable shell, values of the
constants $\lambda'$ and $J'$ and the initial conditions  $R'(0)=
R'_i$, $\dot{R}'(0)=0$. From (\ref{53}) we see that the  condition
$({\cal R'(\tau)}/{\cal R})>1$ is satisfied if $R<R'(\tau)$,
$\forall \ \tau$. In figure(\ref{E}) we show  a sequence $R'(\tau)$
of a damped oscillating solution, obtained through a numerical
integration of (\ref{37}) for $\lambda'= 11.6070, \ \  J'=0.01, \ \
R'_i = 3$.  The internal solution corresponds to the stable static
solution of section 3 ($\Lambda = 0.1$, ${\cal R} = 0.002476$, $J =
0.001$, $R = 1$). We see that the shell oscillates until it reaches
the equilibrium radius $R'_{eq} = 2.7547$. This radius corresponds
to the stable configuration with
$\Lambda'=\Lambda'_{equil}(s(\tau_f), w, R, J')$, where $\tau_f
\geq\ 3\times10^4$, $s (\tau_f) = 1.05335 < 1.53463$. The same
equilibrium configuration is reached for different values of $R'_i$
in a neighborhood of $R'_{eq}$ (we tried $R'_i = 5$, and $2$).

\subsubsection{Generalized Komar energy}
\begin{figure}
\centering \epsfxsize=7.5cm \epsfbox{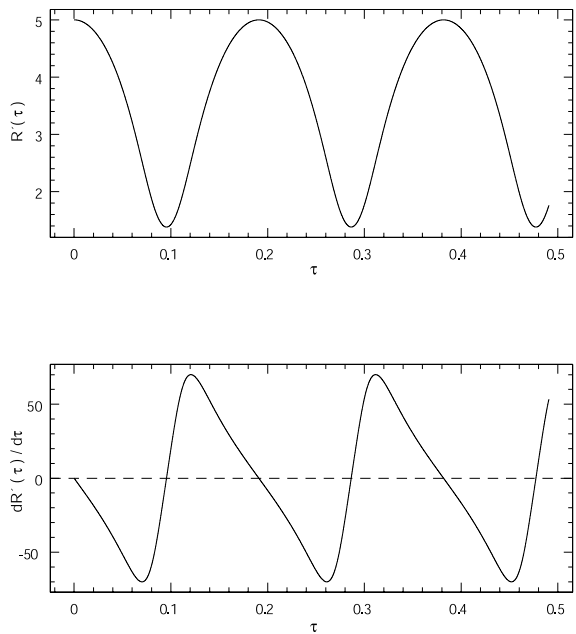} \vspace{-0.3cm}
\caption{\small{The first oscillations of a shell with $R'_i = 5
\,,\; \lambda'= 11.6070, \ \ J'=0.01$. It shows the asymmetry of the
oscillation when it is quite apart from equilibrium.}} \label{F}
\end{figure}

All the knowledge that we have of the external solution is reduced
to $\Sigma'$ as the boundary of $M^+$; i.e, the matter of the
external shell. We may calculate a (time dependent) Komar
mass\cite{komar} per unit of proper length in
$\tilde{M}^{+}=M^{+}\cup \Sigma'$, using the definition
\begin{equation}\label{65}
    M_G=\int_{\bar{\Sigma}}{\xi^b\left(S'^a_b -
    \frac{1}{2}S'\delta^a_b\right)\delta(n)d\bar{\Sigma}_a}\;\;,
\end{equation}
where $\xi^b=k\delta^b_{t^+}\,; \;\bar{\Sigma}: \ t^+=const\,; \
\delta(n)=\frac{1}{\sqrt{g^{rr}}} \ \delta(r-R(\tau))\,;\,
d\bar{\Sigma}_a=\delta^t_a\sqrt{-g}drd\phi dz\,;\\
\sqrt{-g}=\sqrt{g^{rr}}re^{(\gamma^+-\psi^+)}$; $k$ is a constant.
Then (\ref{65}) becomes
\begin{equation}\label{66}
    M_G=k \int_0^{2\pi} \int_0^{e^{{\,-\psi}_{\Sigma'}}} R'(\tau)\left(S'^t_t -
    \frac{1}{2}S'\right)e^{(\gamma^+ - \psi_{\Sigma'})}dz\,
    d\phi\;\;.
\end{equation}
Using $S'^{tt}=\eta'(X^+)^2+p'(h^+)^{tt}$; $(h^+)^{tt} = (g^+)^{tt}
+ (X^+)^{2}$, we have
\begin{equation}\label{67}
    S'^t_t-\frac{1}{2}S'=-(\eta'+p')\left(\frac{1}{2}+\dot{R}'^2e^{2(\gamma'-\psi_{\Sigma'})}  \right)
\end{equation}
choosing $k=-1$, (\ref{66}) becomes:
\begin{equation}\label{68}
    M_G=\Lambda'e^{\gamma'}e^{-\psi_{\Sigma'}}\left(\frac{1+2s^2}{\sqrt{1+s^2}}  \right)  \left(\frac{1}{2}+\dot{R}'^2 e^{2(\gamma'-\psi_{\Sigma'})} \right)
\end{equation}
to obtain $\gamma'$ we have to use the matching condition
\begin{equation}\label{69}
    X^+-X^-=-4\lambda'\sqrt{1+s^2}
\end{equation}
In figure(\ref{F}) we show $R'(\tau)$ and $\dot{R}'$ for $\tau <
0.5$, $R'_i = 5$, and $\lambda'$ and $J'$ equal to the values
corresponding to those of figure (\ref{E}). It is apparent that in
the first instances of the motion $\dot{R}'$ grows faster than it
decreases. It may be inferred, according to order of magnitude
estimates using the reduced quadrupole moment of the shell per unit
of proper length\cite{gravitation}, that the  gravitational
luminosity is larger after  the shell bounces when it reaches  its
minimum radius \cite{piran}. It is clear from (\ref{68}) that most
of the energy radiated away, during the radial oscillation of the
external shell, comes from the term proportional to
$\dot{R}'^2$;i.e, from its kinetic energy as it would be expected.
\subsection{Collapsing and initially expanding solutions}
\begin{figure}
\centering \epsfxsize=7.5cm \epsfbox{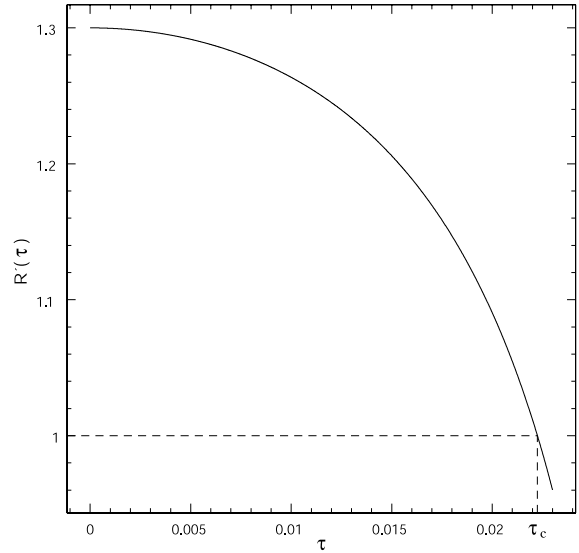}
\vspace{-0.3cm} \caption{\small{A collapsing solution corresponding
to $R'_i = 1.3$ and $R'_{ieq} = 1.3660$; $\tau_c = 0.0222$. The
constant of the model are $w = 0.4038$, $R = 1$, $J = 0.001$, $J' =
0.01$, $\lambda'= 14.59$.}}  \label{G}
\end{figure}

\begin{figure}
\centering \epsfxsize=7.5cm \epsfbox{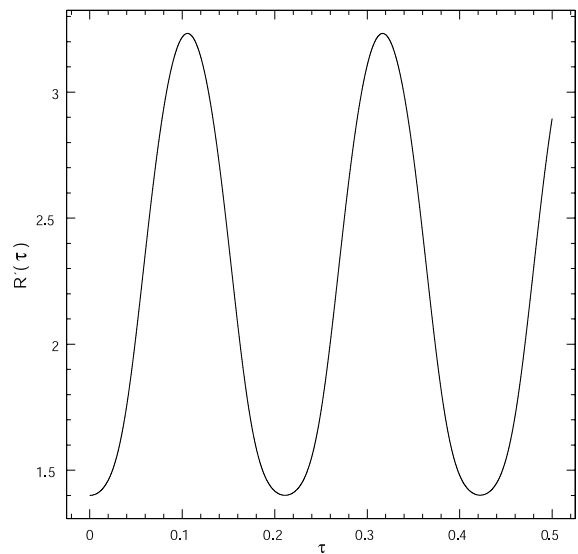}
\vspace{-0.3cm} \caption{\small{An initially expanding solution. The
parameters are the same as in figure(\ref{G}), $R'_i = 1.4$.}}
\label{H}
\end{figure}

An interesting set of solutions of equation (\ref{37}) corresponds
to those with $R_i$ close to an unstable equilibrium configuration
($R'_{ueq}$). For instance, if  $R'_i \, < \,R'_{ueq}$, the solution
of (\ref{37}) is a collapsing shell that  hits the internal shell in
a time of order $\tau_c \sim 10^{-2}$ [see figure (\ref{G})]. On the
other hand for $R'_i\, > \,R'_{ueq}$ the solution is initially
expanding, to settle in a damped oscillating solution with a larger
stable radius. An example is shown in figure (\ref{H}); the unstable
equilibrium radius is $R'_{ueq} = 1.3660$. The shell settles in a
stable  equilibrium radius $R'_f \approx  2.3$.
\section{Final comments}
We have been interested in the relativistic dynamics of cylindrical
shells of counter rotating particles. This system have been
considered before, in some detail, by Apostolatos and
Thorne\cite{AT}. Our main contribution to the knowledge of the model
is the following:
\begin{itemize}
\item We have described the matter composing the shells using the
relativistic kinetic theory adapted to a singular hypersurface. This
approach has the advantage  that the surface energy momentum tensor
satisfies the conservation laws if the distribution function is a
solution of the Liouville's equation.
\item We have analyzed the stability of the equilibrium solutions. We found that the increasing branch
of the plot of equation (\ref{44}) represents the stable
equilibrium, while the decreasing part of the plot are the unstable
equilibrium configurations. This figure was first introduced by
Apostolato and Thorne\cite{AT} for another purpose.
\item The study of the equilibrium solutions was extended to two concentric shells.
 The results are similar as for only one shell [Compare figures (\ref{A}, \ref{B}) with  (\ref{C}, \ref{D})].
\item Armed with the previous results, we analyzed the damped oscillating solutions of the external
 shell in the two shells model. These solutions converge to a
 stable static configuration, depending on the values of the constant
 of the motion and initial conditions.
\item The most interesting result was to show that
the  solutions that start with a radius smaller than the
corresponding radius of a unstable static solution, collapses to hit
the internal shell, whose proper radius may be quite small,
depending on the angular momentum per unit mass of the particles of
the internal shell. The spacetime that results is free of
singularities as long as this angular momentum is not null.
\end{itemize}

Finally, we think that  more information can be obtained from the
equation of motion of a time dependent shell, about the basic
qualitative features of generation of gravitational waves and the
behavior of the matter conforming such shells, in the limit of very
small radius. An interesting  problem would be to relate the motion
of the shell with the C-energy\cite{AT, thorne} carried away by
gravitational waves, while in the interior of the shell we have a
superposition of ingoing and outgoing cylindrical gravitational
waves that satisfies the boundary conditions (\ref{5}). This and
related topics may be discussed elsewhere.

\vspace{1cm}
 \noindent \textbf{\large Acknowledgements}\vspace{.5cm}

The authors VHH and DEB are very grateful to CONICET of Argentina,
and SECyT of the Universidad Nacional de C\'{o}rdoba for financial
support.

\end{document}